\documentclass[prb,letterpaper,aps,floatfix,twocolumn]{revtex4-2}

\usepackage{graphicx}
\usepackage{amsmath,amssymb}
\usepackage{bm}

\usepackage[pdfstartview=FitH,breaklinks=true,bookmarks=true,colorlinks=true,anchorcolor=black,citecolor=red,filecolor=black,menucolor=black,urlcolor=blue,linkcolor=blue]{hyperref}

\begin{document}

\title{Reevaluating Quantum Geometric Criteria for Itinerant Magnetic Instabilities}

\author{Min-Fong Yang}
\email{mfyang@thu.edu.tw}
\affiliation{Department of Applied Physics, Tunghai University, Taichung 40704, Taiwan}

\date{\today}

\begin{abstract}
The interplay between quantum geometry and electron correlation has emerged as a compelling paradigm in quantum many-body physics. Recent studies have highlighted the diagnostic utility of quantum geometry in identifying magnetic instabilities within itinerant electron systems.
In the present work, we critically re-examine these theoretical proposals, which are commonly formulated within the random phase approximation (RPA). To this end, we reformulate the matrix-based RPA instability criterion in the channel representation for generic two-orbital systems, explicitly accounting for multiple channels of magnetic ordering.
Our results demonstrate that magnetic phase transitions are intricately governed by the interplay between the bare susceptibility tensor and the spin interaction matrix. Consequently, prior assertions that instabilities can be predicted solely from the quantum geometric structure of a single-channel susceptibility are valid only under complete channel decoupling in both the interaction and susceptibility matrices.
By adopting the channel representation, our formulation achieves greater physical transparency and computational tractability compared to the conventional orbital-space approach, thereby furnishing a promising alternative for advancing theoretical studies of complex multiorbital systems.
\end{abstract}

\maketitle

\section{INTRODUCTION}

Quantum geometry of Bloch wave functions in momentum space has been recognized as a fundamental concept in condensed matter physics~\cite{Resta1994,Xiao_etal2010}.
Beyond the single-band approximation, Bloch electrons display nontrivial properties arising from their wave functions, thereby highlighting the relevance of quantum geometry.
A representative measure is the Hilbert–Schmidt quantum distance,
\begin{equation}\label{eq:Q_distance}
d_{nm}^2(\mathbf{k},\mathbf{k}+\mathbf{q})=1-|\langle u_n(\mathbf{k})|u_m(\mathbf{k}+\mathbf{q})\rangle|^2 \; ,
\end{equation}
which quantifies the separation between two Bloch states, $|u_n(\mathbf{k})\rangle$ and $|u_m(\mathbf{k}+\mathbf{q})\rangle$, belonging to the $n$-th and $m$-th bands at momenta $\mathbf{k}$ and $\mathbf{k}+\mathbf{q}$, respectively. In close analogy with differential geometry, this quantum distance naturally gives rise to the definition of the quantum geometric tensor, whose real and imaginary parts correspond to the quantum metric and Berry curvature, respectively. These geometric quantities are deeply intertwined with a wide range of phenomena, as evidenced by the extensive studies of quantum geometry in condensed matter physics~\cite{Resta1994,Xiao_etal2010,Rossi2021,%
Torma_etal2022,Torma2023,Liu_etal2025,Yu_etal2025,Gao_etal2025}.

Parallel to these developments, itinerant magnetism has long been a central theme in condensed matter physics, with significant theoretical efforts focused on elucidating its microscopic origins. Motivated by aforementioned progress, recent studies have revisited the stability of ferromagnetic (FM) states from the perspective of quantum geometry~\cite{Wu-DasSarma2020,Kang_etal2024,Kitamura_etal2024,Kitamura_etal2025,%
Kudo-Yanase2025,Oh_etal2025,Hu_etal2025,Shimizu_etal2026,Ding-Claassen2026,%
Kitamura_etal2026}. Notably, it has been shown that a necessary condition for FM ordering in multiband electronic systems can be formulated in terms of the quantum metric of non-interacting states~\cite{Kitamura_etal2024}, thereby establishing a profound link between ferromagnetism and quantum geometry.

This conclusion is based on the result that, within the random phase approximation (RPA) framework, FM ordering in a multiband Hubbard model requires the static bare FM susceptibility $\chi^{(0)}_\mathrm{FM}(\mathbf{q})$ to exhibit a peak at momentum $\mathbf{q}=0$. Specifically,
\begin{equation}
\chi^{(0)}_\mathrm{FM}(\mathbf{q}) = \lim_{\eta \to 0^+}
i \int_{0}^{\infty} dt\, e^{-\eta t} \langle [S(\mathbf{q}, t), S(-\mathbf{q}, 0)] \rangle_0 \; ,
\end{equation}
where $S(\mathbf{q})$ denotes the Fourier transform of the FM magnetization operator. At site $i$, the local magnetization is defined as $S_i = \sum_\sigma \sigma c_{i\sigma}^\dagger c_{i\sigma}$, with $\sigma$ labeling the spin index.
In two dimensions, this condition for $\chi^{(0)}_\mathrm{FM}(\mathbf{q})$ is mathematically equivalent to requiring the curvature tensor,
\begin{equation}
\Omega_{\mu\nu}=\lim_{\mathbf{q}\to 0} \partial_{q_\mu} \partial_{q_\nu} \chi^{(0)}_\mathrm{FM}(\mathbf{q}) \; ,
\end{equation}
to satisfy $\det[\Omega]>0$ and $\mathrm{Tr}[\Omega]<0$.
A valuable insight highlighted by Kitamura~\emph{et al.} in Ref.~\cite{Kitamura_etal2024} is that the static bare FM susceptibility $\chi^{(0)}_\mathrm{FM}(\mathbf{q})$ encodes the information of the quantum distance defined in Eq.~\eqref{eq:Q_distance}. Consequently, the associated curvature tensor $\Omega_{\mu\nu}$ acquires a geometric contribution originating from the quantum geometry of the Bloch wave functions.  Kitamura~\emph{et al.} further demonstrate that the component tied to the quantum metric typically yields a negative contribution to the curvature, thereby enhancing FM fluctuations. This implies that systems can be driven toward an FM instability by tuning the quantum distance and thus the corresponding quantum metric, even while keeping the energy spectrum unchanged~\cite{Kitamura_etal2024,%
Kitamura_etal2025,Kudo-Yanase2025,Oh_etal2025,Shimizu_etal2026}.

Altermagnetism has recently emerged as a distinct symmetry class of collinear magnetic order, characterized by vanishing net magnetization together with momentum-dependent spin splitting~\cite{Hayami2019,Yuan2020,Mazin2021,Smejkal2022-1,Smejkal2022-2}. (For recent reviews, see Refs.~\cite{SongC2025,Jungwirth2025-1,Jungwirth2025-2}.) This unique property renders altermagnets particularly attractive for future spintronics applications~\cite{Bai2024,Guo2025,Jungwirth2025-3}.
Extending the framework established in Ref.~\cite{Kitamura_etal2024}, the quantum-geometric perspective on magnetic instabilities has recently been applied to altermagnetic (AM) ordering~\cite{Heinsdorf2025}. By analyzing the curvature tensor of the static bare AM susceptibility $\chi^{(0)}_\mathrm{AM}(\mathbf{q})$ in non-interacting systems, specifically for collinear AM order defined by the operator $\tilde{S}_i = \sum_\sigma \sigma c_{i\sigma}^\dagger \tau^z c_{i\sigma}$, it has been concluded that a finite quantum metric intrinsically favors AM instabilities over FM ones. Here $\tau^z$ represents the third Pauli matrix acting in orbital space for two-orbital models. This analytical insight is further corroborated by an effective relativistic model of MnTe, which confirms that AM fluctuations peak precisely in momentum regions where the quantum metric is large.

Despite the conceptual appeal of the quantum-geometric perspectives discussed above, conclusions based solely on the behavior of bare susceptibilities in non-interacting systems are inherently incomplete. This is because, beyond the band structure (including the corresponding wavefunctions) and the filling factor encoded in susceptibilities, interaction-induced magnetic instabilities depend critically on the microscopic details of the interactions. In fact, the specific form and strength of these interactions can reconfigure the hierarchy of competing orders, ultimately dictating the prevailing instability. For example, under the repulsive on-site Hubbard–Hund interaction~\cite{Kanamori1963} [see also Eq.~\eqref{eq:H_int}],
it has been shown that, even when the bare susceptibility peaks satisfy $\chi^{(0)}_\mathrm{AM}(\mathbf{q}=0)>\chi^{(0)}_\mathrm{FM}(\mathbf{q}=0)$, the dominant instability can shift from AM to FM as Hund’s coupling increases~\cite{Lu_etal2025,Lu_Hu2026}.

In this work, we employ Ginzburg–Landau theory to elucidate the role of quantum geometry in governing the magnetic instabilities of itinerant electron systems. This framework naturally incorporates various magnetic orderings, facilitating a systematic investigation of their mutual competition. For simplicity, we focus on two-orbital models with on-site interactions, while noting that the extension to more general cases is straightforward. To enhance physical transparency, we characterize the order parameters in terms of distinct channels for each spin component $\sigma$, defined as $\sum_{\alpha,\beta} \langle c^\dag_{i\alpha\sigma} \tau^a_{\alpha\beta} c_{i\beta\sigma}\rangle$, rather than working with the one-body reduced density matrices $\langle c^\dag_{i\alpha\sigma} c_{i\beta\sigma}\rangle$. Here $\tau^a$ ($a = 0, x, y, z$) denote the Pauli matrices acting in orbital space, and $\alpha$, $\beta$ label the orbitals. In this representation, the uniform ordering with $a=0$ corresponds to the FM channel, while $a = x, y, z$ represent distinct AM channels.

The coefficients of the Ginzburg–Landau energy functional are derived within the Hartree–Fock mean-field approximation. From the leading quadratic terms, we find that a magnetic phase transition occurs when the minimal eigenvalue of a stability matrix, constructed from the combination of the interaction-strength matrix and the bare magnetic susceptibility matrix, vanishes. This conclusion is fully consistent with results from multiorbital random phase approximation (RPA) analyses~\cite{Lu_etal2025,Lu_Hu2026,Kubo2007,Graser_etal2009,Altmeyer_etal2016} (see App.~\ref{appendix:int}), reflecting a fundamental equivalence between these two approaches.

Our results place a stringent constraint on recent proposals that formulate instability criteria solely in terms of quantum geometry. Specifically, in the absence of complete channel decoupling, inferring a quantum phase transition from the geometric structure of an isolated susceptibility component is theoretically unjustified. While standard on-site interactions may yield a diagonal interaction matrix in the channel representation, channel decoupling in the bare susceptibility matrix is not guaranteed unless enforced by the symmetries of the non-interacting Hamiltonian. Thus, the reliability of predicting magnetic instabilities through the quantum geometry of an isolated channel hinges strictly on the simultaneous channel decoupling in both the interaction and susceptibility matrices. When this condition fails, a comprehensive analysis of the full stability matrix, which incorporates all inter-channel mixing effects, becomes indispensable.

Furthermore, even under complete channel decoupling, the concavity of single-channel susceptibilities alone is insufficient to predict the ground state. Instead, the relative peak magnitudes across competing channels must be assessed to identify the prevailing magnetic order. To illustrate, we analyze the two-orbital model with repulsive on-site Hubbard–Hund interactions discussed in Ref.~\cite{Lu_etal2025}, where both the interaction and susceptibility matrices are diagonal in the channel representation. In this setting, channel decoupling disentangles the instability conditions into independent Stoner-like criteria for distinct magnetic sectors. In alignment with Ref.~\cite{Lu_etal2025}, we show that the dominant instability can shift from AM to FM as the Hund’s coupling $J_H$ is tuned. Notably, this transition occurs even when the bare AM susceptibility exceeds its FM counterpart, since the $J_H$-driven enhancement of the effective FM interaction eventually outweighs the initial bias in the bare response. This result underscores that the prevailing instability is dictated by the interplay between the bare magnetic response and the effective interaction strength, rather than relying solely on the quantum geometric features of a single channel.

The remainder of this paper is organized as follows. In Sec.~II, we formulate the Ginzburg–Landau energy functional and derive the free energy within the Hartree–Fock mean-field approximation, through which our matrix-based criterion for magnetic instabilities is established. Within this framework, we elucidate the fundamental limitations of previous criteria that rely exclusively on the quantum geometric structure of an isolated susceptibility channel. In Sec.~III, we apply this formalism to a specific two-orbital model to investigate the competition between FM and AM instabilities, explicitly demonstrating how the prevailing phase is dictated by the interplay between bare susceptibilities and interaction strengths. Finally, we conclude our paper in Sec.~IV. The derivation of the interaction matrix in the channel representation is described in App.~\ref{appendix:int} for completeness.

\section{framework of Ginzburg-Landau theory}\label{sec:GL}

In this section, we derive the Ginzburg–Landau free energy within the Hartree–Fock mean-field approximation to investigate the onset of magnetic instabilities. Near a continuous phase transition, where the order parameters are infinitesimally small, the free energy can be systematically expanded in powers of these parameters and retain only the lowest-order terms. The coefficients of leading quadratic terms are intrinsically linked to the  magnetic susceptibility, which measures how strongly the system responds to an infinitesimal tendency toward magnetic ordering. To illustrate the formalism concretely, we focus on a representative two-orbital model with on-site interactions. This choice ensures physical transparency while the generalization to more complex multi-orbital systems with finite-range interactions is straightforward.

Assuming static collinear magnetic orderings, the local order parameters at site $i$ are defined as
\begin{equation}
m_i^a = \sum_\sigma \sigma \langle c_{i\sigma}^\dagger \tau^a c_{i\sigma} \rangle \; ,
\end{equation}
where $\tau^a$ ($a = 0, x, y, z$) are the Pauli matrices acting in the orbital space. Within the Hartree–Fock approximation, the interaction Hamiltonian becomes
\begin{align}
&H_\mathrm{int} \approx -\sum_{i,a} \Delta_i^a \left( \sum_{\sigma} \sigma c_{i\sigma}^\dagger \tau^a c_{i\sigma} \right) + E_0 \; , \\
&E_0 = \frac{1}{4} \sum_i\sum_{a,b} \, [U^{(s)}]_{ab} \, m_i^a \, m_i^b \; .
\end{align}
Here $\Delta_i^a=(1/2)\sum_b \, [U^{(s)}]_{ab} \, m_i^b$ represents the effective field associated with the $a$-th magnetic channel, and $[U^{(s)}]_{ab}$ is the spin interaction matrix that couples channels $a$ and $b$ (refer to App.~\ref{appendix:int} for the explicit form of this matrix). Upon diagonalizing the resulting quadratic mean-field Hamiltonian, we obtain the energy spectrum $E_{\alpha,\sigma}$. The free energy is then given by
\begin{equation}
F = E_0 - T \sum_{\alpha,\sigma} \ln \left( 1 + e^{-\beta E_{\alpha,\sigma}} \right) \equiv E_0 + \tilde{F} \; .
\end{equation}

In the vicinity of the phase transition, where the order parameters are infinitesimally small ($m_i^a \simeq 0$ and thus $\Delta_i^a \simeq 0$), the Ginzburg–Landau free energy functional can be systematically constructed by expanding $F$ in a power series of $m_i^a$. Utilizing the thermodynamic identity
\begin{equation}
\frac{\partial \tilde{F}}{\partial \Delta_i^a} = - \sum_\sigma \sigma \langle c_{i\sigma}^\dagger \tau^a c_{i\sigma} \rangle = -m_i^a \; ,
\end{equation}
we establish the link between the free energy and the order parameters. Within the paramagnetic phase, linear response theory gives that the induced order parameter $m_i^a$ is proportional to the effective field $\Delta_j^b$ via the non-interacting (bare) susceptibility
\begin{equation}
m_i^a \simeq 2 \sum_{j,b}\, [\chi^{(0)}]^{ab}_{i,j} \, \Delta_j^b \; .
\end{equation}
Consistent with this linear relationship, the bare susceptibility $[\chi^{(0)}]^{ab}_{i,j}$ of the non-interacting Hamiltonian is defined by the second-order variation of the free energy evaluated at $\Delta_i^a = 0$,
\begin{equation}\label{eq:bare_sus}
[\chi^{(0)}]^{ab}_{i,j} = -\frac{1}{2} \frac{\partial^2 \tilde{F}}{\partial\Delta_i^a \partial\Delta_j^b} \; .
\end{equation}
Therefore, up to quadratic order in $m_i^a$, the Ginzburg-Landau free energy reads
\begin{align}\label{eq:G-L}
F &\simeq \frac{1}{4} \sum_{i,j} \, [U^{(s)}]_{cb} \, m_i^b \left( \delta_{a,c}\,\delta_{i,j} - [\chi^{(0)}]^{cd}_{i,j} \, [U^{(s)}]_{da} \right) m_j^a \nonumber \\
&= \frac{1}{4} \sum_\mathbf{q} \, [U^{(s)}]_{cb} \, m_\mathbf{-q}^b \left( I - \chi^{(0)}(\mathbf{q})\; U^{(s)}\right)_{ca} m_\mathbf{q}^a  \; .
\end{align}
Here the summation convention for channel indices is assumed. In deriving the final expression, translational symmetry of the non-interacting system is invoked, with $m_\mathbf{q}^a$ defined as the Fourier transform of $m_i^a$.

According to the Ginzburg-Landau theory, a phase transition occurs when the determinant of the second-order coefficient matrix vanishes. From Eq.~\eqref{eq:G-L}, it is evident that magnetic instability arises when the minimal eigenvalue of the $4\times4$ stability matrix, $\mathbb{M}(\mathbf{q})=I-\chi^{(0)}(\mathbf{q})\,U^{(s)}$, becomes zero. This criterion is consistent with results obtained from the multiorbital random phase approximation (RPA)~\cite{Lu_etal2025,Lu_Hu2026,Kubo2007,Graser_etal2009,%
Altmeyer_etal2016,note_RPA}. It demonstrates that the onset of instabilities is generally governed in a nontrivial manner by both the interaction strengths and the matrix elements of the bare susceptibility. A simplification arises only when both the interaction matrix $U^{(s)}$ and the bare susceptibility matrix $\chi^{(0)}(\mathbf{q})$ become diagonal. In such a case, the instability condition reduces to the Stoner form, $1-[\chi^{(0)}(\mathbf{q})]^{aa}\,[U^{(s)}]_{aa}=0$ (no sum on $a$), for a given channel $a$. In this situation, a peak of $[\chi^{(0)}(\mathbf{q})]^{aa}$ at the ordering momentum can signal instability if its product with $[U^{(s)}]_{aa}$ is indeed the largest value among all channels. However, when the matrices cannot be reduced to diagonal forms, it becomes theoretically untenable to infer instability directly from the behavior (and thus the quantum geometric structure) of $[\chi^{(0)}(\mathbf{q})]^{aa}$. Under such circumstances, a comprehensive analysis of the whole stability matrix $\mathbb{M}(\mathbf{q})$, which incorporates all inter-channel mixing effects, becomes indispensable.

In certain interaction Hamiltonians widely adopted in the literature, the interaction matrices $[U^{(s)}]_{ab}$ are diagonal within the channel representation, despite their non-diagonal form in the original orbital basis. This underscores that, beyond offering physical transparency, the channel representation facilitates representational conciseness. A concrete example is the on-site Hubbard–Hund interaction~\cite{Kanamori1963}, where the interaction matrix $[\mathcal{U}^{(s)}]_{ab}$ indeed adopts a diagonal form when expressed in the channel representation (see App.~\ref{appendix:int} for the explicit derivation). Nevertheless, the diagonal nature of the interaction matrix \emph{per se} is insufficient to justify a single-channel analysis. To rigorously employ the quantum geometric structure of a single component $[\chi^{(0)}(\mathbf{q})]^{aa}$ as a criterion for quantum phase transitions, complete channel decoupling within the bare susceptibility matrix is strictly requisite. Such decoupling is not a generic feature but is fundamentally dictated by the specific symmetries of the non-interacting Hamiltonian.

We emphasize that, even with complete channel decoupling, the concavity of a single-channel susceptibility is insufficient to uniquely determine the ground state. To illustrate this point, in the following section we analyze the two-orbital model investigated in Ref.~\cite{Lu_etal2025}, where both the interaction and susceptibility matrices are diagonal within the channel representation. Consistent with Ref.~\cite{Lu_etal2025}, we show that the dominant magnetic instability arises from the combined influence of bare susceptibilities and effective interaction strengths, rather than being dictated solely by single-channel susceptibility behavior.

\section{competition between FM and AM instabilities}\label{sec:AM-FM}

In the following, we examine the two-orbital model with the Hamiltonian given by
\begin{equation}\label{eq:2band}
\mathcal{H}_0(\mathbf{k})= [\varepsilon_0(\mathbf{k})-\mu] \,\tau^0 + \varepsilon_x(\mathbf{k})\,\tau^x + \varepsilon_z(\mathbf{k})\,\tau^z \; ,
\end{equation}
where $\mu$ is the chemical potential and $\tau^{0,x,z}$ are the identity and Pauli matrices acting in the orbital space. The corresponding energy eigenvalues are given by
\begin{equation}
E_\pm(\mathbf{k}) = \varepsilon_0(\mathbf{k}) \pm \sqrt{\varepsilon_x^2(\mathbf{k}) + \varepsilon_z^2(\mathbf{k})} -\mu
\end{equation}
with the associated eigenstates parameterized by a mixing angle $\theta_\mathbf{k}$,
\begin{equation}
|u_+(\mathbf{k})\rangle = \begin{pmatrix} \cos\frac{\theta_\mathbf{k}}{2} \\ \sin\frac{\theta_\mathbf{k}}{2} \end{pmatrix}, \quad %
|u_-(\mathbf{k})\rangle = \begin{pmatrix} -\sin\frac{\theta_\mathbf{k}}{2} \\ \cos\frac{\theta_\mathbf{k}}{2} \end{pmatrix} \nonumber \; .
\end{equation}
The angle $\theta_\mathbf{k}$ is uniquely determined by the relations,
\begin{equation}\label{eq:angles}
\sin\theta_\mathbf{k}=\frac{\varepsilon_x(\mathbf{k})} {\sqrt{\varepsilon_x^2(\mathbf{k}) + \varepsilon_z^2(\mathbf{k})}} \; , \quad \cos\theta_\mathbf{k}=\frac{\varepsilon_z(\mathbf{k})}
{\sqrt{\varepsilon_x^2(\mathbf{k}) + \varepsilon_z^2(\mathbf{k})}} \; .
\end{equation}

In the band basis, the static bare susceptibility matrix among channels reads
\begin{align}
&[\chi^{(0)}(\mathbf{q})]^{ab} =
-\sum_{\mathbf{k}}\sum_{n,n'=\pm} \langle u_n(\mathbf{k}) | \tau^a | u_{n'}(\mathbf{k}+\mathbf{q}) \rangle \nonumber \\
&\qquad\times\langle u_{n'}(\mathbf{k}+\mathbf{q}) | \tau^b | u_n(\mathbf{k}) \rangle
\,\frac{f[E_n(\mathbf{k})] - f[E_{n'}(\mathbf{k}+\mathbf{q})]}{E_n(\mathbf{k}) - E_{n'}(\mathbf{k}+\mathbf{q})} \; .
\label{eq:bare_sus_q}
\end{align}
Following the insights of Refs.~\cite{Kitamura_etal2024,Kitamura_etal2025,%
Kudo-Yanase2025,Oh_etal2025,Shimizu_etal2026}, each diagonal element of the susceptibility matrix $[\chi^{(0)}(\mathbf{q})]^{aa}$ encodes information regarding a ``generalized'' quantum distance [cf. Eq.~\eqref{eq:Q_distance} ]:
\begin{equation}\label{eq:gen_Q_distance}
\tilde{d}_{nm}^2(\mathbf{k},\mathbf{k}+\mathbf{q})=1-|\langle \tilde{u}_n(\mathbf{k})|u_m(\mathbf{k}+\mathbf{q})\rangle|^2 \; ,
\end{equation}
where $|\tilde{u}_n(\mathbf{k})\rangle \equiv \exp(i\frac{\pi}{2}\tau^a)|u_n(\mathbf{k})\rangle$ denotes a transformed (``twisted") Bloch state. This measure quantifies the geometric separation between the states $|\tilde{u}_n(\mathbf{k})\rangle$ and $|u_m(\mathbf{k}+\mathbf{q})\rangle$. Consequently, the associated curvature tensor $\Omega_{\mu\nu}$ incorporates a geometric contribution arising from the quantum geometry of the Bloch wavefunctions under the operation of the generator $\tau^a$. Notably, this quantum-geometric perspective is generally restricted to the diagonal elements and does not readily extend to the off-diagonal components of the susceptibility matrix~\cite{note_QG}. The applicability of quantum geometry is therefore constrained to specific systems where inter-channel fluctuations are decoupled. In more generic settings, however, the intricate interplay among channels obscures any straightforward geometric interpretation.

Furthermore, even in the case of complete channel decoupling, the true instability cannot be reliably inferred solely from the susceptibility of a single channel. This situation can happen if competing instabilities arise across different sectors, and thus a comparative analysis of all relevant channels becomes necessary.

To illustrate these points, we adopt the model discussed in Ref.~\cite{Lu_etal2025} to investigate the specific competition between FM and AM instabilities. From Eq.~\eqref{eq:bare_sus_q}, the static bare susceptibility matrix $[\chi^{(0)}(\mathbf{0})]^{ab}$ at momentum $\mathbf{q}=\mathbf{0}$ becomes
\begin{widetext} 
\begin{align}
[\chi^{(0)}(\mathbf{0})]^{ab} =& -\sum_{\mathbf{k}}\sum_{n=\pm} \langle u_n(\mathbf{k}) | \tau^a | u_n(\mathbf{k}) \rangle \langle u_n(\mathbf{k}) | \tau^b | u_n(\mathbf{k}) \rangle\, f'[E_n(\mathbf{k})] \nonumber \\
&-\sum_{\mathbf{k}}\sum_{\substack{n,n'=\pm\\n'\neq n}} \langle u_n(\mathbf{k}) | \tau^a | u_{n'}(\mathbf{k}) \rangle \langle u_{n'}(\mathbf{k}) | \tau^b | u_n(\mathbf{k}) \rangle\, \frac{f[E_n(\mathbf{k})] - f[E_{n'}(\mathbf{k})]}{E_n(\mathbf{k}) - E_{n'}(\mathbf{k})}  \; .
\label{eq:bare_sus_0}
\end{align}
\end{widetext} 
The first term represents the intraband contribution arising from states near the Fermi surface, while the second term accounts for the interband contribution involving transitions between different bands.

For the model employed in Ref.~\cite{Lu_etal2025}, the dispersion components $\varepsilon_a(\mathbf{k})$ in Eq.~\eqref{eq:2band} are given by
\begin{align}\label{eq:dispersions}
\begin{aligned}
\varepsilon_0(\mathbf{k}) = & -2t_0[\cos(k_x)+\cos(k_y)] \\
&- 2t_1[\cos(2k_x)+\cos(2k_y)] \, , \\
\varepsilon_x(\mathbf{k}) =\;& 4t_2\sin(k_x)\sin(k_y) \, , \\
\varepsilon_z(\mathbf{k}) =\;& 2t_0^\prime[\cos(k_x)-\cos(k_y)] \\
&+ 2t_1^\prime[\cos(2k_x)-\cos(2k_y)] \; .
\end{aligned}
\end{align}
Here $\{t_0, t_1\}$ and $\{t_0', t_1', t_2\}$ represent the orbital-independent and orbital-dependent hopping parameters, respectively. Based on these expressions, we observe that $\sin\theta_\mathbf{k}$ defined in Eq.~\eqref{eq:angles} transforms as an odd function under reflection across either the $x$ or $y$ axis (i.e., $k_x \rightarrow -k_x$ or $k_y \rightarrow -k_y$). In contrast, $\cos\theta_\mathbf{k}$ is antisymmetric under reflection across the diagonal line $y=x$ (i.e., $k_x \leftrightarrow k_y$).
By evaluating the matrix elements $\langle u_{n'}(\mathbf{k}) | \tau^a | u_n(\mathbf{k}) \rangle$ and performing the summation over the Brillouin zone, we find that the static bare susceptibility matrix in the channel representation exhibits a diagonal form:
\begin{equation}
    [\chi^{(0)}(\mathbf{0})]^{ab} =
    \begin{pmatrix}
    \chi^{(0)}_{00} & 0 & 0 & 0 \\
    0 & \chi^{(0)}_{xx} & 0 & 0 \\
    0 & 0 & \chi^{(0)}_{yy} & 0 \\
    0 & 0 & 0 & \chi^{(0)}_{zz}
    \end{pmatrix}\, .
\end{equation}
Here the vanishing of all off-diagonal elements during the momentum-space integration is fundamentally attributed to the symmetry properties of $\sin\theta_\mathbf{k}$ and $\cos\theta_\mathbf{k}$.

Consequently, the fluctuations of distinct channels decouple for the model under consideration. The explicit expressions for the diagonal elements, representing FM, $\tau^x$-type AM, $\tau^y$-type AM, and $\tau^z$-type AM fluctuations, respectively, are given by
\begin{align}
\chi^{(0)}_{00} &= -\sum_{\mathbf{k}}\sum_{n=\pm} f'[E_n(\mathbf{k})] \; , \\
\chi^{(0)}_{xx} &= \chi^{(0)}_{00} - \sum_{\mathbf{k}}\cos^2\theta_\mathbf{k} \,\Bigg\{ 2\,\frac{f[E_+(\mathbf{k})] - f[E_-(\mathbf{k})]}{E_+(\mathbf{k}) - E_-(\mathbf{k})} \nonumber \\
&\hspace{2cm} - \sum_{n=\pm} f'[E_n(\mathbf{k})] \Bigg\} \; ,\\
\chi^{(0)}_{yy} &= -2\,\sum_{\mathbf{k}}\frac{f[E_+(\mathbf{k})] - f[E_-(\mathbf{k})]}{E_+(\mathbf{k}) - E_-(\mathbf{k})} \; , \\
\chi^{(0)}_{zz} &= \chi^{(0)}_{00} - \sum_{\mathbf{k}}\sin^2\theta_\mathbf{k} \,\Bigg\{ 2\,\frac{f[E_+(\mathbf{k})] - f[E_-(\mathbf{k})]}{E_+(\mathbf{k}) - E_-(\mathbf{k})} \nonumber \\
&\hspace{2cm} - \sum_{n=\pm} f'[E_n(\mathbf{k})] \Bigg\} \; .
\end{align}
We note that the expressions for $\chi^{(0)}_{00}$ and $\chi^{(0)}_{zz}$ coincide with those derived in Sec. III of Ref.~\cite{Roig_etal2024}.

In contrast to the purely intraband FM channel $\chi^{(0)}_{00}$, the AM susceptibilities $\chi^{(0)}_{xx}$ and $\chi^{(0)}_{zz}$ comprise both intraband and interband contributions. The competition between these two processes determines the leading instability: altermagnetism is stabilized when the interband contribution predominates, whereas ferromagnetism prevails if the intraband term is dominant. Notably, for the $\tau^z$-type AM instability, a non-vanishing $\varepsilon_x(\mathbf{k})$ term in Eq.~\eqref{eq:2band} (implying $\sin^2\theta_\mathbf{k} > 0$) is indispensable; otherwise, the FM and $\tau^z$-type AM instabilities become degenerate. Similarly, the realization of the $\tau^x$-type AM instability necessitates a finite $\varepsilon_z(\mathbf{k})$ term (implying $\cos^2\theta_\mathbf{k} > 0$) to lift the degeneracy between the FM and $\tau^x$-type AM sectors.

Under the on-site Hubbard–Hund interaction~\cite{Kanamori1963}, the interaction matrix becomes diagonal in the channel indices, as shown in App.~\ref{appendix:int}. From the results established in Sec. II, the criteria for magnetic instabilities in each channel are respectively determined by
\begin{align}\label{eq:criteria}
\begin{aligned}
&\frac{1}{2}(U+J_{H})\,\chi^{(0)}_{00}=1 \, , \\
&\frac{1}{2}(U-J_{H})\,\chi^{(0)}_{xx}=1 \, , \\
&\frac{1}{2}(V-J')\,\chi^{(0)}_{yy}=1 \, ,  \\
&\frac{1}{2}(U-J_{H})\,\chi^{(0)}_{zz}=1 \, .
\end{aligned}
\end{align}
Here $U$ denotes the intra-orbital Hubbard repulsion, $V$ the inter-orbital density interaction, $J_H$ the Hund's exchange coupling, and $J'$ the pair-hopping term. Based on these conditions, it is evident that increasing the Hund's coupling $J_H$ can drive a transition from an AM-type instability to an FM-type one. This occurs because a larger $J_H$ simultaneously enhances the effective interaction in the FM channel while suppressing it in the AM channels. Consequently, the FM instability can prevail even in regimes where the bare AM susceptibilities ($\chi^{(0)}_{xx}$ or $\chi^{(0)}_{zz}$) exceed the FM one ($\chi^{(0)}_{00}$). This conclusion is in excellent agreement with the findings reported in Ref.~\cite{Lu_etal2025}.

This case clearly demonstrates that, rather than relying solely on the behavior (such as the concavity) of single-channel susceptibilities, one must account for the quantitative interplay between their peak magnitudes and the explicit structure of the interaction matrix. It is this combined effect of bare response and interaction strength that ultimately dictates the dominant instability among competing channels.

Furthermore, we note that the off-diagonal elements of the static bare susceptibility matrix in the channel representation can become finite once the dispersion components $\varepsilon_a(\mathbf{k})$ are modified. For example, replacing $\varepsilon_x(\mathbf{k})$ in Eq.~\eqref{eq:dispersions} by $\varepsilon_x(\mathbf{k}) = 4t_2\sin(k_x)\sin(k_y) + t_2^\prime[\cos(k_x)-\cos(k_y)]$ renders the off-diagonal susceptibilities $\chi^{(0)}_{xz}$ and $\chi^{(0)}_{zx}$ nonzero, thereby coupling the $\tau^x$-type and $\tau^z$-type AM fluctuations. The static bare susceptibility matrix at $\mathbf{q}=0$ then takes the form
\begin{equation}
    [\chi^{(0)}(\mathbf{0})]^{ab} =
    \begin{pmatrix}
    \chi^{(0)}_{00} & 0 & 0 & 0 \\
    0 & \chi^{(0)}_{xx} & 0 & \chi^{(0)}_{xz} \\
    0 & 0 & \chi^{(0)}_{yy} & 0 \\
    0 & \chi^{(0)}_{zx} & 0 & \chi^{(0)}_{zz}
    \end{pmatrix}\, .
\end{equation}
The emergence of finite $\chi^{(0)}_{xz}$ and $\chi^{(0)}_{zx}$ can be traced to the change in the symmetry properties of the product $\sin\theta_\mathbf{k}\cos\theta_\mathbf{k}$. After the above modification, this quantity is no longer odd under either $k_x \rightarrow -k_x$, $k_y \rightarrow -k_y$, or the diagonal reflection $k_x \leftrightarrow k_y$. Consequently, the symmetry cancellation in the momentum-space integration is lost, leading to nonvanishing values of $\chi^{(0)}_{xz}$ and $\chi^{(0)}_{zx}$.

As a consequence, the instability criterion in Eq.~\eqref{eq:criteria} for AM fluctuations in the $\tau^x$-$\tau^z$ space is replaced by
\begin{equation}
\frac{\chi^{(0)}_{xx}+\chi^{(0)}_{zz}}{2} \pm \sqrt{\left(\frac{\chi^{(0)}_{xx}-\chi^{(0)}_{zz}}{2}\right)^2 + \left(\chi^{(0)}_{xz}\right)^2}  = \frac{2}{U-J_H} \, . \nonumber
\end{equation}
The instability is therefore determined by the eigenvalues of the full susceptibility matrix rather than by any individual channel susceptibility. This explicitly demonstrates that, in the absence of complete channel decoupling, the instability depends on a nontrivial interplay between diagonal and off-diagonal susceptibility components. Consequently, the quantum geometry associated with a single channel can no longer be regarded as a direct indicator of the instability, even though the spin-interaction matrix remains diagonal in the channel representation.

\section{conclusions}

In this work, we provide a critical re-evaluation of recent theoretical proposals that invoke quantum geometry to predict magnetic instabilities in itinerant electron systems. Previous investigations have suggested that, through its intrinsic connection to the concavity of single-channel bare susceptibilities, the quantum metric of non-interacting states can elucidate the emergence of either FM or AM instabilities. However, we point out that such a geometric perspective is fundamentally incomplete. Its validity is strictly contingent upon the decoupling of channels in both electron interactions and fluctuations, a condition that is not always satisfied in generic multi-band systems.

To systematically substantiate this claim, we construct a rigorous matrix-based criterion for magnetic instabilities, derived from a Ginzburg-Landau functional within the Hartree-Fock mean-field approximation for a generic two-orbital system. Specifically, a magnetic phase transition out of the paramagnetic state occurs when the minimum eigenvalue of the stability matrix, $\mathbb{M}(\mathbf{q}) = I - \chi^{(0)}(\mathbf{q})\, U^{(s)}$, vanishes. This matrix incorporates the interplay between the bare susceptibility tensor $\chi^{(0)}(\mathbf{q})$ and the spin interaction matrix $U^{(s)}$, demonstrating that instabilities are governed by the coupled dynamics of multiple channels in a non-trivial manner. Consequently, analyzing quantum phase transitions via a single-channel susceptibility component (and its quantum geometric structure) is valid only if the interaction matrix and the bare susceptibility matrix are both diagonal in the channel representation.

We note that our established criterion for magnetic instabilities is consistent with results obtained via the standard multiorbital RPA formalism~\cite{Kubo2007,Graser_etal2009,Altmeyer_etal2016}. Nevertheless, our derivation shows that the channel representation offers a more physically transparent and computationally convenient framework for analyzing multiorbital systems than the conventional orbital-space approach. This methodological observation provides a valuable alternative for future theoretical investigations into complex multiorbital materials.

\begin{acknowledgments}
The author would like to thank Yu-Wen Lee for enlightening discussions. This research was supported by Grant No. NSTC 114-2112-M-029-001 and NSTC 114-2112-M-029-006 of the National Science and Technology Council of Taiwan.
\end{acknowledgments}

\appendix
\section{interaction matrices in channel representation} \label{appendix:int}

In the following discussions, we restrict our attention to on-site interactions within a two-orbital model. For notational conciseness, we suppress the site dependence of all physical quantities and omit the summations over site indices.

Assuming collinear order, the Hartree-Fock mean-field decoupling leads to
\begin{align}
H_\mathrm{int} \approx &\; \langle c_{l_1 \uparrow}^\dagger c_{l_2 \uparrow} \rangle \, [\bar{U}_{\uparrow\uparrow}]_{l_3 l_4}^{l_1 l_2} \, c_{l_3 \uparrow}^\dagger c_{l_4 \uparrow} \nonumber \\
& + \langle c_{l_1 \uparrow}^\dagger c_{l_2 \uparrow} \rangle \, [\bar{U}_{\uparrow\downarrow}]_{l_3 l_4}^{l_1 l_2} \, c_{l_3 \downarrow}^\dagger c_{l_4 \downarrow} \nonumber \\
& + \langle c_{l_1 \downarrow}^\dagger c_{l_2 \downarrow} \rangle  \, [\bar{U}_{\uparrow\downarrow}]_{l_3 l_4}^{l_1 l_2} \, c_{l_3 \uparrow}^\dagger c_{l_4 \uparrow} \nonumber \\
& + \langle c_{l_1 \downarrow}^\dagger c_{l_2 \downarrow} \rangle \, [\bar{U}_{\uparrow\uparrow}]_{l_3 l_4}^{l_1 l_2} \, c_{l_3 \downarrow}^\dagger c_{l_4 \downarrow} \; .
\end{align}
Here spin-inversion symmetry in $H_\mathrm{int}$ is assumed and the summation convention is adopted throughout. Henceforth, symbols for the interaction and susceptibility matrices equipped with a bar (e.g., $\bar{U}$, $\bar{\chi}^{(0)}$) denote quantities expressed in the orbital representation. Conversely, their unbarred counterparts represent the corresponding quantities defined within the channel representation.

Define the particle–hole operators in the charge and spin sectors as $n_{l_1 l_2} = c_{l_1 \uparrow}^\dagger c_{l_2 \uparrow} + c_{l_1 \downarrow}^\dagger c_{l_2 \downarrow}$ and $m_{l_1 l_2} = c_{l_1 \uparrow}^\dagger c_{l_2 \uparrow} - c_{l_1 \downarrow}^\dagger c_{l_2 \downarrow}$, the mean-field interaction Hamiltonian thus becomes
\begin{align}\label{eq:H_MF}
H_\mathrm{int}^\mathrm{MF} = \frac{1}{2} \langle n_{l_1 l_2} \rangle \, [\bar{U}^{(c)}]_{l_3 l_4}^{l_1 l_2} \, n_{l_3 l_4} 
- \frac{1}{2} \langle m_{l_1 l_2} \rangle \, [\bar{U}^{(s)}]_{l_3 l_4}^{l_1 l_2} \, m_{l_3 l_4} \; .
\end{align}
Here $\bar{U}^{(c)} = \bar{U}_{\uparrow\downarrow} + \bar{U}_{\uparrow\uparrow}$ and $\bar{U}^{(s)} = \bar{U}_{\uparrow\downarrow} - \bar{U}_{\uparrow\uparrow}$ represent the charge and the spin interaction matrices, respectively.

To make the physical meaning more transparent, the particle–hole operators with orbital indices can be transformed into those for distinct channels, $n^a = [\tau^a]_{l_1 l_2} n_{l_1 l_2}$ and $m^a = [\tau^a]_{l_1 l_2} m_{l_1 l_2}$, where $\tau^a$ with $a=0, x, y, z$ denote the Pauli matrices acting in the orbital space. Notice that the Pauli matrices satisfy a completeness relation (also known as the Fierz identity~\cite{Okun_book1982}), which can be expressed using their Hermitian property:
\begin{equation}\label{eq:Fierz_id}
\frac{1}{2} \sum_{a} \, [\tau^{a}]_{l_{1}l_{2}} \; [\tau^{a}]^*_{l_{4}l_{3}}=\delta_{l_{1}l_{4}} \, \delta_{l_{2}l_{3}}  \, ,
\end{equation}
where $[\tau^{a}]^*$ denotes the complex conjugate of the Pauli matrix. By inserting this identity into Eq.~\eqref{eq:H_MF} on both the left- and right-hand sides of $[\bar{U}^{(c),(s)}]_{l_3 l_4}^{l_1 l_2}$, the mean-field interaction Hamiltonian can be rewritten as
\begin{equation}
H_\mathrm{int}^\mathrm{MF} = \frac{1}{2} \langle n^a \, \rangle [U^{(c)}]_{ab} \, n^b - \frac{1}{2} \langle m^a \, \rangle [U^{(s)}]_{ab} \, m^b \; ,
\end{equation}
where the transformed interaction matrices in the channel representation are given by
\begin{equation}\label{eq:U_transform}
[U^{(c),(s)}]_{ab} = \frac{1}{4} \, [\tau^a]^*_{l_1 l_2} \, [\bar{U}^{(c),(s)}]_{l_3 l_4}^{l_1 l_2} \, [\tau^b]^*_{l_3 l_4} \; .
\end{equation}

Interestingly, the same conclusion can be reached within the matrix RPA framework. The bare susceptibility matrix $[\chi^{(0)}(\mathbf{q})]^{ab}$ in the channel representation is defined as
\begin{equation}
[\chi^{(0)}(\mathbf{q})]^{ab} = [\tau^{a}]_{l_{1}l_{2}} \; [\tau^{b}]_{l_{3}l_{4}} \; [\overline{\chi}^{(0)}(\mathbf{q})]_{l_{3}l_{4}}^{l_{1}l_{2}} \, ,
\end{equation}
where $[\overline{\chi}^{(0)}(\mathbf{q})]_{l_{3}l_{4}}^{l_{1}l_{2}}$ is the bare susceptibility matrix in the orbital space. By inserting the identity of Eq.~\eqref{eq:Fierz_id} between the matrices $[\overline{\chi}^{(0)}(\mathbf{q})]_{l_{3}l_{4}}^{l_{1}l_{2}}$ and $[\bar{U}^{(s)}]_{l_{3}l_{4}}^{l_{1}l_{2}}$ in each term of the matrix RPA perturbation series~\cite{Kubo2007,Graser_etal2009,Altmeyer_etal2016},
\begin{align}
\overline{\chi}_\mathrm{spin}^\mathrm{RPA}(\mathbf{q}) %
&= [I - \overline{\chi}^{(0)}(\mathbf{q}) \; \bar{U}^{(s)}]^{-1} \; \overline{\chi}^{(0)}(\mathbf{q}) \nonumber \\
& = \overline{\chi}^{(0)}(\mathbf{q}) + \overline{\chi}^{(0)}(\mathbf{q}) \; \bar{U}^{(s)} \; \overline{\chi}^{(0)}(\mathbf{q}) + \cdots \, ,
\end{align}
the RPA susceptibility in terms of $[\chi^{(0)}(\mathbf{q})]^{ab}$ can be expressed as
\begin{align}
\chi_\mathrm{spin}^\mathrm{RPA}(\mathbf{q}) %
& = \chi^{(0)}(\mathbf{q}) + \chi^{(0)}(\mathbf{q}) \; U^{(s)} \; \chi^{(0)}(\mathbf{q}) + \cdots \nonumber \\
&= [I - \chi^{(0)}(\mathbf{q}) \; U^{(s)}]^{-1} \; \chi^{(0)}(\mathbf{q}) \, .
\end{align}
Here the corresponding $4\times4$ spin interaction matrices $[U^{(s)}]_{ab}$ in the channel representation is again given by Eq.~\eqref{eq:U_transform}.

In general, the interaction matrices in the orbital space are not diagonal, whereas their channel representation can be. The analysis can thus be simplified by working with $[U^{(c),(s)}]_{ab}$, rather than the original orbital form. This occurs for the on-site Hubbard–Hund interaction~\cite{Kanamori1963},
\begin{align}
H_\mathrm{int}&=
U \sum_{\mathbf{i},l=x,y} n_{\mathbf{i},l,\uparrow} n_{\mathbf{i},l,\downarrow}
+V \sum_{\mathbf{i}} \sum_{s,s^\prime} n_{\mathbf{i},x,s} n_{\mathbf{i},y,s^\prime}
\nonumber \\%
&+J_H \sum_{\mathbf{i}} \sum_{s,s^\prime} c_{\mathbf{i},x,s}^\dag c_{\mathbf{i},y,s^\prime}^\dag c_{\mathbf{i},x,s^\prime} c_{\mathbf{i},y,s}
\nonumber \\%
&+J^\prime \sum_{\mathbf{i}} c_{\mathbf{i},x,\uparrow}^\dag c_{\mathbf{i},x,\downarrow}^\dag c_{\mathbf{i},y,\downarrow} c_{\mathbf{i},y,\uparrow} + h.c. \, . \label{eq:H_int}
\end{align}
Here $U$ denotes the intra-orbital Hubbard repulsion, $V$ the inter-orbital density interaction, $J_H$ the Hund's exchange coupling, and $J^\prime$ the pair-hopping term. The corresponding spin interaction matrix with indices $(l_{1}l_{2}) = (xx, yy, xy, yx)$ reads~\cite{Lu_etal2025,Lu_Hu2026}
\begin{equation}
    [U^{(s)}]_{l_{3}l_{4}}^{l_{1}l_{2}} =
    \begin{pmatrix}
    U & J_{H} & 0 & 0 \\
    J_{H} & U & 0 & 0 \\
    0 & 0 & J' & V \\
    0 & 0 & V & J'
    \end{pmatrix}\, .
\end{equation}
By using Eq.~\eqref{eq:U_transform}, we find that the corresponding channel representation becomes diagonal, $[U^{(s)}]_{ab}=g_a\,\delta_{a,b}$ with the diagonal matrix elements $g_a$ for distinct channels read
\begin{equation*}
(g_0, g_x, g_y, g_z) = \left(\frac{U + J_H}{2}, \frac{V + J'}{2}, \frac{V - J'}{2}, \frac{U - J_H}{2} \right) \; .
\end{equation*}
For systems respecting rotation symmetry, the interaction parameters are constrained by the identity $U = V + J' + J_H$. In this case, we have $g_x=g_z=(U-J_H)/2$.




\begin{thebibliography}{99}
\bibitem{note_RPA}
Although $\chi^{(0)}(\mathbf{q})$ and $U^{(s)}$ are Hermitian, their product is generally non-Hermitian unless they commute. Numerical calculations therefore employ either the generalized eigenvalue problem or a symmetrization method to obtain the eigenvalues of the stability matrix, $\mathbb{M}(\mathbf{q})=I-\chi^{(0)}(\mathbf{q})\,U^{(s)}$.
\end{thebibliography}
\end{document}